# On the equivalence of the Nernst theorem and its consequence

Shanhe Su[+], Yinghui Zhou[+], Guozhen Su, Jincan Chen[*]

Department of Physics, Xiamen University, Xiamen 361005, People's Republic of China

One general consequence of the Nernst theorem is derived, i.e., the various heat capacities of a thermodynamic system under different constraints approach zero as the temperature approaches absolute zero. The temperature dependence of the heat capacity of any thermodynamic system at ultra-low temperatures is revealed through this consequence. Moreover, the general form and the simplest expression of the heat capacities of thermodynamic systems at ultra-low temperatures are deduced. Some significant discussion and results are given. One new research method is provided by using this consequence. Finally, the equivalence between the Nernst theorem and its consequence is rigorously proved, so that this consequence may be referred to another description of the third law of thermodynamics.

---

[+]The two authors contributed equally to this work

[*]Email: jcchen@xmu.edu.cn

The Nernst theorem is one main description of the third law of thermodynamics [1-4]. It is well known that at absolute zero temperature, the many properties of thermodynamic systems may be directly derived from Nernst theorem. For example, the heat capacity $C_v$ at constant volume, heat capacity $C_P$ at constant pressure, coefficient of thermal expansion $\alpha$, and thermal pressure coefficient $\beta$ of thermodynamic systems [2, 4-7] trend to zero as the temperature approaches absolute zero. Yan and Chen [8] further discussed the relation between the Nernst theorem and the heat capacity $C_y$ when all the generalized coordinates $y$ remain unchanged. So far the problem of the general relation between the various heat capacities of a thermodynamic system under different constraints and the Nernst theorem has been rarely discussed, but it is an interesting problem worth studying in thermodynamic theory.

For a general thermodynamic system, the fundamental equation of thermodynamics is given by

$$dU = TdS + \sum_{i=1}^{n} Y_i dy_i, \tag{1}$$

where $U$ and $S$ are the internal energy and entropy of the system, $y_i$ and $Y_i$ are the generalized coordinates and corresponding generalized forces, $n$ is the number of generalized coordinates, and $T$ is the absolute temperature. Using Eq. (1), one gets

$$C_y = T(\partial S / \partial T)_y. \tag{2}$$

Using the Legendre transformation $L = U - \sum_{j=1}^{l} y_j Y_j$, we obtain the differential expression of the new thermodynamic function $L$ as

$$dL = TdS + \sum_{i=l+1}^{n} Y_i dy_i - \sum_{j=1}^{l} y_j dY_j. \tag{3}$$

From Eq. (3), one obtains

$$C_x = T(\partial S / \partial T)_x, \tag{4}$$

where $x$ represents $(n-l)$ generalized coordinates and $l$ generalized forces. When $l=0$, $x$ represents all generalized coordinates $y$ and the third term on the right-hand side in Eq. (3) equals zero. When $l=n$, $x$ represents all generalized forces $Y$ and the second term on the right-hand side in Eq. (3) equals zero. $C_x$ is the heat capacity when $x$ remains unchanged. Thus, $C_x$ includes $C_y$ as well as $C_Y$, where $C_Y$ is the heat capacity when $Y$ is invariant. For example, the heat capacity at constant volume $C_v$ and the heat capacity at constant pressure $C_P$ are two simplest forms of $C_x$. Thus, $C_x$ includes the various heat capacities of a thermodynamic system under different constraints.

From Eqs. (1) and (3), one obtains

$$\left(\frac{\partial S}{\partial U}\right)_y = \frac{1}{T} = \left(\frac{\partial S}{\partial L}\right)_x, \tag{5}$$

$$\left(\frac{\partial^2 S}{\partial U^2}\right)_y = -\frac{1}{T^2 C_y}, \tag{6}$$

and

$$\left(\frac{\partial^2 S}{\partial L^2}\right)_x = -\frac{1}{T^2 C_x}. \tag{7}$$

Eq. (6) indicates that in the regime of $T>0$, $C_y>0$ since $(\partial^2 S/\partial U^2)_y < 0$ is the equilibrium stability condition of a general thermodynamic system [1, 4, 8, 9]. As pointed out in Refs. [10-13], the condition of stability of a thermodynamic system is the concavity of the entropy [14]. It is clearly seen from Eq. (5) that in the regime of $T>0$, both the $L \sim S$ curve and the $U \sim S$ curve have the same curved shape and the slope of each point on the curve equals $1/T$. It implies the fact that the second partial derivatives of the entropy $S$ with respect to $U$ and $L$ have the same sign, i.e., $(\partial^2 S/\partial U^2)_y<0$ and $(\partial^2 S/\partial L^2)_x<0$. Thus, it requires that in the regime of $T>0$,

$$C_x>0. \tag{8}$$

According to the Nernst theorem that the entropy change associated with any isothermal reversible process of a condensed system approaches zero as the temperature approaches absolute zero [4], i.e.,

$$\lim_{T\to 0}(\Delta S)_T=0, \tag{9}$$

and another thermodynamic relation $M=L-TS$, we can obtain

$$dM=-SdT+\sum_{i=l+1}^{n}Y_i dy_i-\sum_{j=1}^{l}y_j dY_j \tag{10}$$

and

$$\lim_{T\to 0}S=-\lim_{T\to 0}\left(\frac{\partial M}{\partial T}\right)_x=-\lim_{T\to 0}\frac{M-L}{T}=-[\lim_{T\to 0}\left(\frac{\partial M}{\partial T}\right)_x-\lim_{T\to 0}\left(\frac{\partial L}{\partial T}\right)_x]. \tag{11}$$

From Eqs. (3), (4), and (11), one obtains

$$\lim_{T\to 0}\left(\frac{\partial L}{\partial T}\right)_x=\lim_{T\to 0}T\left(\frac{\partial S}{\partial T}\right)_x=\lim_{T\to 0}C_x=0. \tag{12}$$

This shows clearly that Eq. (12) is a direct consequence of the Nernst theorem and a universal result of thermodynamics. Obviously, $\lim_{T\to 0}C_y=0$ is included in Eq. (12), so that the results in Refs. [8, 15] can be directly derived from this Letter.

Eq. (12) enormously enriches the contents of thermodynamics. At ultra-low temperatures, if the heat capacity does not satisfy Eq. (12), its corresponding equation of state is incompatible with the third law of thermodynamics and cannot be used to discuss the properties of the system at ultra-low temperatures.

Through the measurement of the heat capacity of a thermodynamic system and the comparison with Eq. (12), it can be judged whether the equation of state of the system is compatible with the third law of thermodynamics and may be used to discuss the properties of the system at ultra-low temperatures. This means that through Eq. (12), it can establish a new research method to discuss the physical properties of a system not only at the absolute zero of temperature but also at ultra-low temperatures. For example, for magnetic materials satisfying Curie's equation, the heat capacity $C_H$ at isomagnetic intensity may not satisfy Eq. (12), and consequently, the third law of thermodynamics requires Curie's equation to fail at ultra-low temperatures [4, 6]. For classical gases, all the heat capacities do not satisfy Eq. (12), and consequently, the equations of state of classical gases are not compatible with the third law of thermodynamics [16] and can not be used to discuss the properties of gases at ultra-low temperatures. For various systems whose heat capacity may be expressed as

$$C_x = aT^\lambda, \tag{13}$$

where $a > 0$ and $\lambda \leq 0$, the equations of state of these systems are not compatible with the third law of thermodynamics because Eq. (13) does not satisfy Eq. (12). This shows that the heat capacity of any thermodynamic system at ultra-low temperatures does not appear in the form of Eq. (13).

For a system consisting of $N$ magnetic dipoles, where each dipole has a magnetic moment $\mu$ and a choice of two orientations with the corresponding energies $-\varepsilon$ and $+\varepsilon$, the mean magnetic moment and the heat capacity at isomagnetic intensity of the system are, respectively, given by [17]

$$M = N\mu_B \tanh \frac{\varepsilon}{kT} \tag{14}$$

and

$$C_H = Nk\left(\frac{\varepsilon}{kT}\right)^2 \operatorname{sech}^2 \frac{\varepsilon}{kT}, \tag{15}$$

where $\mu_B$ is the Bohr magneton and $k$ is the Boltzmann constant. Eq. (15) satisfies Eq. (12), and consequently, Eq. (14) is compatible with the third law of thermodynamics and can be used to discuss the properties of magnetic materials at ultra-low temperatures. For superflow $^3He$, $C_v \propto T^3 \ln T$; superconductor, $C_v \propto e^{-\Delta/T}$; and superfluid Fermi gas, $C_v \propto T^{3/2} e^{-\Delta/T}$ [7, 18]. The heat capacities of these systems satisfy Eq. (12) and the corresponding equations of state are also compatible with the third law of thermodynamics. When the heat capacity of a thermodynamic system is given by

$$C_x = A_1 T^{\chi} + A_2 T^{\delta} + A_3 T^{\gamma} + ..., \tag{16}$$

where $\chi > 0$, $\chi < \delta < \gamma$, $A_i \geq 0$ (i=1,2,3,...), but $A_i (i=1,2,3,...)$ cannot equal zero at the same time, the equation of state of the system is compatible with the third law of thermodynamics and can be used to discuss the properties of the system at ultra-low temperatures because Eq. (16) satisfies Eq. (12). It can be seen from Eq. (16) that the simplest form satisfying Eq. (12) is given by

$$C_x = AT^{\chi}, \tag{17}$$

where $A$ is a finite value larger than zero, which depends on the concrete properties of a thermodynamic system but is independent of temperature. For example, quantum Bose gas, $C_v \propto T^{3/2}$; quantum Fermi gas, $C_v \propto T$; photon gas, $C_v \propto T^3$; electron gas in metal, $C_v \propto T$; and superflow $^4He$, $C_v \propto T^3$ [6, 14, 17, 18]. These heat capacities satisfy Eq. (17), and consequently, the corresponding equations of state are compatible with the third law of thermodynamics and can be used to discuss the properties of the systems described above at ultra-low temperatures.

Eq. (4) can be used to derive the entropy of a general thermodynamic system as [4, 6]

$$S = S(T,x) = S(T_0,x) + \int_{T_0}^{T} (C_x/T)dT . \tag{18}$$

When $T_0 = 0$, Eq. (18) can be rewritten as

$$S(T,x) = S(0,x) + \int_{0}^{T} (C_x/T)dT . \tag{19}$$

Mathematically, there are no additional requirements for Eq. (19). Physically, some additional requirements are necessary for Eq. (19), because a thermodynamic system always obeys the Nernst theorem, which requires that Eq. (12) must be satisfied and the integral term $\int_{0}^{T}(C_x/T)dT$ in Eq. (19) is a finite value. In other words, the heat capacitors $C_x$ of a real thermodynamic system can ensure the integral term $\int_{0}^{T}(C_x/T)dT$ in Eq. (19) to be a finite value.

Now we continue to prove that the Nernst theorem can be derived from Eq. (12). According to Eq. (12), we can prove that for a real thermodynamic system,

$$0 < \int_{0}^{T_j} [C_x(x,T)/T]dT \equiv F(x,T_j) . \tag{20}$$

For a thermodynamic system evolving along a quasistatic reversible adiabatic process, $x$ changes from $x'$ to $x''$, and $T$ changes from $T_1$ to $T_2$, whereas the entropy $S$ remains unchanged. Using Eqs. (19) and (20), we can obtain

$$S(0,x'') - S(0,x') = F(x',T_1) - F(x'',T_2) . \tag{21}$$

It is seen from Eq. (21) that if $S(0,x'') > S(0,x')$ is assumed, one can find a positive $T_1$ to ensure that $F(x',T_1)$ is small enough and $S(0,x'') - S(0,x') > F(x',T_1)$. In such a case, it requires that $F(x'',T_2)$ must be negative, which is in contradiction with Eq. (20); similarly, if $S(0,x'') < S(0,x')$ is assumed, one can find a positive $T_2$ to ensure that $F(x'',T_2)$ is small enough and $S(0,x') - S(0,x'') > F(x'',T_2)$, and consequently, it requires that $F(x',T_1)$ must be negative, which is also in contradiction with Eq. (20). Thus, we must have a conclusion that $S(0,x") = S(0,x')$ as $T \to 0$, which is accurately the Nernst theorem.

So far it has been proved that both the Nernst theorem and Eq. (12) may be mutually deducible and they are equivalent. Thus, like the Nernst theorem, Eq. (12) may be also taken as a description of the third law of thermodynamics, i.e., the various heat capacities of a thermodynamic system under different constraints approach zero as the temperature approaches absolute zero. Such a description may be simply referred as the heat capacity description of the third law of thermodynamics.

**Acknowledgements**

This work is supported by the National Natural Science Foundation (No. 12075197), People's Republic of China.

Press)